# Mining changes of user expectations over time from online reviews

Tianjun HOU, corresponding author, tianjun.hou@centralesupelec.fr, Laboratoire Genie Industriel, CentraleSupélec, Université Paris-Saclay, Gif-sur-Yvette, France;

Bernard YANNOU, bernard.yannou@centralesupelec.fr, Laboratoire Genie Industriel, CentraleSupélec, Université Paris-Saclay, Gif-sur-Yvette, France;

Yann LEROY, yann.leroy@centralesupelec.fr, Laboratoire Genie Industriel, CentraleSupélec, Université Paris-Saclay, Gif-sur-Yvette, France;

Emilie POIRSON, emilie.poirson@ec-nantes.fr, LS2N, Ecole Centrale de Nantes, Nantes, France

**Abstract**

Customers post online reviews at any time. With the timestamp of online reviews, they can be regarded as a flow of information. With this characteristic, designers can capture the changes in customer feedback to help set up product improvement strategies. Here we propose an approach for capturing changes of user expectation on product affordances based on the online reviews for two generations of products. First, the approach uses a rule-based natural language processing method to automatically identify and structure product affordances from review text. Then, inspired by the Kano model which classifies preferences of product attributes in five categories, conjoint analysis is used to quantitatively categorize the structured affordances. Finally, changes of user expectation can be found by applying the conjoint analysis on the online reviews posted for two successive generations of products. A case study based on the online reviews of Kindle e-readers downloaded from amazon.com shows that designers can use our proposed approach to evaluate their product improvement strategies for previous products and develop new product improvement strategies for future products.

## 1. Introduction

Online product reviews have become a viable and valuable source for collecting user requirements and preference for product development, especially for those designers who need to continually refresh their products in a competitive marketplace [1-3]. Compared with traditional user requirement identification methods such as focus-group exercises, interviews, the large amount of readily accessible online review data enables designers to identify customer needs in a timely and efficient manner [4]. As these online reviews get updated in real-time, designers can constantly acquire new feedback at all times. This unprecedented characteristic provides designers new ways to gain knowledge on the market structure and competitive landscape to support their decisions. The research reported here provides an approach for capturing changes in user expectation using online reviews. The approach can be used to evaluate and develop product improvement strategies.

The start point of our research is a discussion on the definition of user requirements. Previous scholarship has implicitly assumed that user requirements concerned mainly product features, i.e. product components, product attributes, etc. Natural language processing algorithms have been proposed to identify the words and expressions related to product features from online review sentences.

However, we find that this assumption is problematic. Product features alone cannot cover all the significant issues addressed in customer reviews [5]. Reviewers describe not only their judgment on product features, but also their experiences of using the product, including how they use it and in what context, etc. For example, in a 5-star review of the Kindle Paperwhite 3, the reviewer said: "I can read books at night without hurting my eyes". The product feature mentioned in this sentence would be that of screen brightness. However, it cannot possibly be identified by natural language processing algorithm because it is not clearly written.

To tackle this issue, we introduce the concepts of product affordance and usage context in online review analysis. These concepts have been widely used in design science to describe potential user–product behaviors [6, 7]. By observing the linguistic patterns, a rule-based natural language processing method is proposed to automatically identify and structure product affordances, usage contexts and their associated perceptual words from online reviews.

Next, inspired by the Kano model, we introduce conjoint analysis to quantitatively categorize the structured affordances into the six categories of the Kano model – also considering the *questionable* category which is often omitted. Specifically, we focus on the affordances on which people have opposite perceptions. For example, for an e-reader, the perception of some reviewers is that it is *easy* to carry with the hands, while others reported that it is *hard* to carry with the hands. The weight of each perception in the star-rating is quantified using conjoint analysis based on an ordered logit model. The Kano model is then innovatively applied to explain the results of the conjoint analysis. Finally, by applying the proposed method on the online reviews posted in different time-spans, the changes of user expectation can be captured.

This research contributes to domain literature on two fronts. First, inspired by the Kano model, we provide an innovative way to use the conjoint analysis in the affordance-based design and online review analysis. The Kano model has traditionally been used for categorizing product functions and attributes, but here we extend its use to categorizing product affordances. Conjoint analysis has traditionally been a survey-based statistical technique, but here we extend its use to online review analysis. This study thus injects fresh relevancy into the affordance-based design and conjoint analysis to serve for today's trends of big data analytics.

Second, unlike previous research in online review analysis, our approach relies on the unprecedented but understudied characteristic of online review data, i.e. online reviews are indeed an information flow. With the traditional requirement identification methods like questionnaires, focus groups, it is difficult to revert to user expectation information obtained at a given point in the past. Therefore, compared with the traditional methods, our approach brings added-value knowledge to support decisions in product design, i.e. the possibility of tracing patterns of change in user expectation.

The remainder of this paper is organized as follows. Section 2 describes the related work. Section 3 presents the research framework. Section 4 describes the rule-based natural language processing method used to automatically identify product affordances, usage contexts and the associated user perceptions. Section 5 describes the conjoint analysis method employed to categorize the affordances in the Kano model. Section 6 describes the case study based on the online reviews of Kindle e-readers posted from 2013 to 2018. The dynamic changes in user expectation for multiple affordances of e-reader are analyzed in order to forge product improvement strategies. Section 7 concludes the research.

## 2. Literature review
### 2.1 From unstructured data to structured data

One major difference between online review data and the data provided by traditional user requirement identification methods is the shift from structured transactional data to unstructured user-generated content [8]. The words and expressions that are meaningful to product design must be identified and structured before providing further insights into decision-making.

Online review structuring is dominated by feature-based opinion mining [8, 9], which entails summarizing reviewers' sentiment orientations towards product features from each in-review sentence. Various methods have been proposed to make use of the extracted product features and sentiment orientations to gain insights for product design. Liu, Jin, Ji, Harding and Fung [10] filtered helpful reviews serving design based on the frequency of product features mentioned in the review and the strength of the sentiment. Tuarob and Tucker [11] identified lead users from social media data based on the frequency of unexpected product features mentioned by the reviewers. Tuarob and Tucker [12] used social media data to quantify product favorability based on the sentiment strength and orientations. Jin, Ji and Gu [13] analyzed product strengths and weaknesses based on the comparative opinions of product features. Zhang, Sekhari, Ouzrout and Bouras [14] proposed several improvement strategies based on the strength of negative sentiment for each product feature. Qi, Zhang, Jeon and Zhou [15] sorted product features based on their influence on the strength and the polarity of sentiment.

However, product features alone do not cover all aspects of what users like and dislike in the online reviews [5], and so researchers have turned attention to other aspects of user requirement. De Weck, Ross and Rhodes [16] mapped co-occurrences using product abilities (or 'ilities') identified from product technical literature to learn the relations among these abilities. Product ability was defined as desired properties of a system, such as flexibility, maintainability, etc., which often manifest themselves after the system has been put to initial use. Shu, Srivastava, Chou and Lai [17] conducted an explorative study on co-occurrences of cue phrases, like "as opposed to", "notice", etc., and novel usages. Novel usage was defined as usage that was not intended by designers when designing the product.

### 2.2 The role of natural language processing in online review analysis

Ample research in computer science has highlighted the viability of natural language processing in the automatic transaction from unstructured text data to structured data. Methods proposed in previous studies can be collapsed into two groups: the rule-based method and the supervised machine learning method [14]. The rule-based method identifies meaningful words and expressions using several IF … THEN … statements. The hypothesis part, i.e. IF …, mainly concerns manually constructed regular patterns of linguistic features, such as the part-of-speech, the grammatical dependency, the lemma, etc., and statistical features, such as the frequency of occurrence, the probability of co-occurrence, etc. These patterns generally came from domain knowledge or observation. For example, in previous work, one of the commonly agreed rules used in feature-based opinion mining method was that product features are described with noun phrases that appear repeatedly.

Unlike rule-based methods, supervised machine learning methods do not rely on manually constructed regular patterns. Instead, they require a mass of high-quality manually-structured data to train probabilistic human language models, such as the Hidden Markov Model (HMM)[18], conditional random fields [13], etc. The trained models are then used to automatically structure text data. However, this kind of method carries the disadvantages of being domain-dependent [14, 19]. New training data are needed when supervised machine learning methods are applied to the reviews of new product categories. Preparing the corpora is a challenge because creating a large-scaled annotated corpus can be very expensive [19].

### 2.3 Online reviews as an information flow

People post online reviews at any time. Therefore, the online review data are always renewing. They can be considered as information flow. Based on this characteristic, it is possible to capture changes in data by comparing the current data against the data in the past, which is why the computation of dated review data holds so much promise.

Tuarob and Tucker [4] attempted to predict product market adoption by analyzing the correlation degree of correlation between product longevity and product sales using online social media data in a series of time-spans. Product longevity was defined based on the number of positive statements and negative statements in social media data. Suryadi and Kim [20] found that frequency of occurrence of different product features has different influences on sales rank. Online reviews could thus be used to highlight the product features that have the biggest influence on sales rank. Zhang, Sekhari, Ouzrout and Bouras [14] analyzed the correlation between the strength of sentiment of each product feature and product sales, and used the correlation to devise a method for target product features that need to be improved. Min, Yun and Geum [21] studied the dynamic change in the number of positive reviews and negative reviews on mobile applications over time. They used the Kano model to explain the dynamic patterns of change.

Previous scholarship has mainly focused on what trends can be concluded by analyzing the correlation between frequency of occurrence of product features and the product's sales, but without providing the drivers behind these trends, i.e. how user expectation evolves over time.

### 2.4 Affordance-based design

The concept of affordance was introduced into product design to address a gap in traditional functional modeling, i.e. some products cannot be represented by input/output models of function [6]. For example, a chair for sitting on does not involve any transformation of material, information or energy, and yet sit-ability is surely an affordance that should be considered when designing the chair.

As is mentioned in the research of Maier and Fadel [22], the scope of affordance can be very broad. All kinds of potential behaviors that can happen between the product and another system (e.g., the user) can be regarded as affordance. Once the concept of affordance was introduced into product design [23], it brought energetic discussions on how to describe the difference between function and affordance [24, 25]. The debate goes on, but the consensus is that affordances do not include the notion of teleology [24]. Compared with function, affordance emphasizes the potentiality of the behaviors between two systems that would not be possible with either system in isolation [6]. That is why affordance has commonly been described in the form of verb-ability, such as maintainability, upgradability, sit-ability, etc. [26]. It includes not only the abilities endowed by designers, but also the misuses and innovative uses made by end-users.

### 2.5 The Kano model

The Kano model is a seminal theory for product development and customer satisfaction (Figure 1) [27]. It classifies product features into five attribute categories based on the correlation between customer preferences and the quality or intensity of the feature:

1) *Must-be* attributes, which consist of the basic product criteria. Customers will be extremely dissatisfied if these basic criteria are not fulfilled, although fulfillment will not increase satisfaction level because customers take their presence for granted.
2) *Performance* attributes, which when present increase satisfaction levels but when absent decreases satisfaction

levels proportionally. This type of attribute provides customer loyalty for firms.
3) *Attractive* or *must-have* or *exciter* attributes, which usually act as a weapon to differentiate companies from their competitors because their functional presence generates absolutely positive satisfaction whereas customers will not be dissatisfied at all without it.
4) *Indifferent* attributes, which make little contribution to customer satisfaction regardless of whether they are present or absent in a product.
5) *Reverse* attributes, which should be removed from a product because their functional presence is actually detrimental to customer satisfaction.

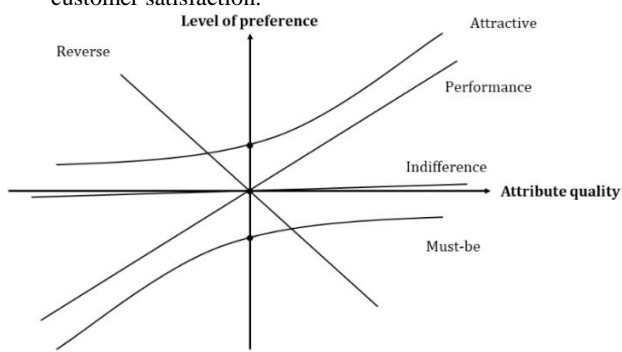

Figure 1 Mapping attributes to the Kano model [27]

To do so, a Kano survey is used to ascertain the customer satisfaction classification of an attribute (Figure 2). During the survey, customers are asked pairs of questions. For each attribute, each participant is asked to rate their satisfaction level if 1) the attribute is present on the product, and 2) the attribute is absent on the product. Then, a Kano evaluation matrix is constructed based on the survey results. Finally, for each attribute, the designers count the number of participants for each category in the Kano model, and the count number can determine one or several dominant categories. Note that the table in Figure 2 includes a new category value of "questionable". This is not an actual Kano model category. Rather, answers that fit this criterion usually signify that the questions were phrased incorrectly, that the customer did not understand the question, or that there was a mistake in selecting a survey answer.

Figure 2 Kano survey questions and the Kano evaluation matrix

## 2.6 Conjoint analysis

Conjoint analysis is a survey-based statistical technique used in market research that helps determine how people value different attributes that make up an individual product or service [28]. The objective of the conjoint analysis is to determine what combination of a limited number of attributes have the strongest influence on respondent choices or decision-making [29]. A controlled set of potential products or services is shown to survey respondents, and by analyzing their different preference levels to these products, the implicit valuation of the individual elements making up the product or service can be determined. These implicit valuations can be used to create market models that estimate market share, revenue, and even the profitability of new design [30].

## 3. Research framework

Figure 3. schematizes our research framework. As current feature-based opinion mining methods provide only limited information [5], our approach here adopts the concepts of affordance and usage context in online review analysis. In the first part of our research, a rule-based natural language processing method is proposed to extract product affordances, usage contexts and the associated perceptual words from the unstructured online review text. This is the basic portion of our research, which includes affordance description formalization, linguistic features identification, identification rules construction and implementation.

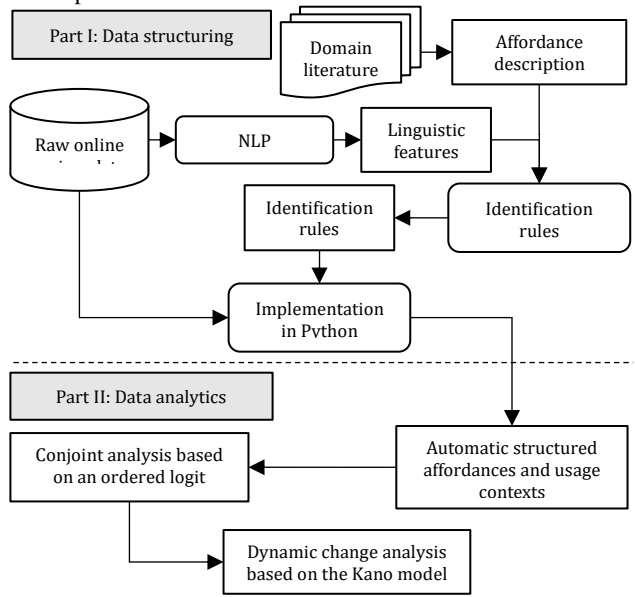

Figure 3 Research framework

The value of online review data added to product design depends on one of its less-studied characteristics: online reviews can be regarded as information flow. Therefore, in the second part, we analyze the changes in user expectation through the online reviews of Kindle e-readers posted from 2013 to 2018. This portion includes conjoint analysis, dynamic change analysis with the Kano model and improvement strategy investigation.

## 4. Extracting product affordances and usage contexts

### 4.1 Formalizing the affordance description

We use the following affordance description form to structure affordances in our study:

*Afford the ability to [action word] [action receiver] [perceived quality] [usage context]*

This description form is derived from the three basic affordance description forms summarized by Hu and Fadel [26]: "verb-ability", "verb noun-ability", "transitive verb noun/intransitive verb". In the basic description forms, the indispensable element is the verb, namely *action word* in our proposed form, which defines the potential behavior between the product and another system (e.g. end user, postman). Alternative elements are the object of the verb, namely *action receiver* in our proposed form, which further defines the receiver of the behavior, and the suffix -ability, which shows that affordance is indeed a kind of potentiality (Section 2.4). Two alternative elements, namely *perceived quality* and *usage context*, are added in the proposed form in order to capture more detailed information related to the product affordances. Perceived quality defines in which dimension and how well the product can support a potential behavior in the reviewer's view [6]. Usage context defines the physical surroundings in which the behaviors take place, such as geographical location, weather, etc. For example, the ability to read books *at night*. Specifying the usage context enables designers to easily target the determining features of product affordance. For example, obviously, the determining features are different for *the ability to read books in the dark* and *the ability to read books in bright sunlight*.

### 4.2 Constructing identification rules

One of the characteristics of our method is that it is capable of processing all kinds of product categories. According to Section 2.2, supervised machine learning methods require new training data each time the product category is changed [19]. Therefore, our method does not rely on learning techniques on a particular product category, but on general heuristics-based regular linguistic or statistical patterns. This section describes the rules that we built to identify the four elements in the affordance description form: action word, action receiver, perceived quality and usage context.

Action words are targeted first, as they are indispensable elements of the description form. Alternative elements are then identified based on the identification of action words. Based on our previous work [31] and our analysis in Section 4.1, we use the following identification rules:

Identification of action word:
- IF the word $w$ is a verb, THEN $w$ is labeled as an action word
- IF $w$ is a noun or an adjective AND $w$ has the suffix -ility, -ilities, -able AND $w$ is derived from a verb, THEN $w$ is labeled as an action word
- IF the word $w$ is a stative verb or an emotional verb, THEN $w$ is not labeled as an action word

Identification of action receiver:
- IF the word $w$ is an object of its head word $h$, AND $h$ is an action word, THEN $w$ is labeled as an action receiver.
- IF the word $w$ is an action word in the clausal modifier of its head word $h$, AND $w$ has its own subject AND $h$ is a noun, THEN $h$ is labeled as an action receiver.

Identification of perceived quality:
- IF the word $w$ is an adverb AND its head word $h$ is an action word in verb or adjective AND $w$ has an antonym, THEN $w$ is labeled as a perceived quality.
- IF the word $w$ is an adjective AND its head word $h$ is an action word in noun AND $w$ has an antonym, THEN $w$ is labeled as a perceived quality.
- IF the word $w$ is an adjective AND it is the open clausal complement of its head word $h$, AND $h$ is an action word, THEN $w$ is labeled as a perceived quality.
- IF the word $w$ is the negation of its head word $h$, AND $h$ is an action word, THEN $w$ is labeled as a perceived quality.

Identification of usage context:
- IF the word $w$ is a positional preposition AND $w$ is the head word of $h$ AND $h$ is an object of the preposition of $w$, THEN $w$ is labeled as a usage context.

### 4.3 Implementing the proposed rules with natural language processing programs

The rules we proposed in Section 4.2 enable us to identify product affordances and usage contexts through linguistic features (the words underlined in the identification rules under Section 4.2). To summarize, the following linguistic features are needed: 1) the part-of-speech, to show whether a word is adjective, noun, verb, preposition, etc.; 2) the grammatical dependency relation, to navigate in the dependency tree and show the grammatical structure of the sentence, such as object, subject, etc. In particular, a head word means the parent of the word in dependency tree; 3) the word derivation, to show the original form of the word; 4) the verb category, to show whether a verb is an emotional verb or a stative verb; 5) the antonym. The first three linguistic features are provided by many open-source NLP packages offering part-of-speech-tagging, parsing and lemmatizing algorithms, such as NLTK[1], Stanford CoreNLP[2], or Spacy[3]. The information on the verb category and the antonym is provided by WordNet. WordNet is a large relational lexical database of English, including antonymy relation. The builder of WordNet has categorized verbs into fourteen groups, including an emotional verb group and a stative verb group[4].

## 5. Capturing changes in user expectation

Inspired by the Kano model, this section investigates changes of user expectation based on the structured affordance descriptions, conjoint analysis. The analysis begins with a discussion on the definition of user preference and perception.

### 5.1 Clarifying the definition of user preference and perception

Previous feature-based sentiment analysis has generally confused the concept of preference with the concept of perception. The scholarship had implicitly assumed that the perceptual words associated with product features indicated whether customers liked or disliked it. Studies used sentiment lexicon to determine the polarity of the sentiment expressed through perceptual words [8, 14].

However, we find that this assumption is a gross approximation. Preference refers to whether the customer likes or dislikes the product, while perception refers to the way in which the product is regarded, understood or interpreted [32-35]. For example, the word *low* in "low battery capacity" is considered a derogatory term in many sentiment lexicons such as Vader[5], SentiWordNet[6], DAL[7], but it does not necessarily mean that the customer disliked the battery. A customer who is used to carrying a power bank can tolerate this feature. Actually, it is commonplace to see the people posting online reviews have different perceptions on the same affordance, and people having the same perception can nevertheless give different star-ratings. For example, for the affordance "*ability to read book*" offered by an e-reader, the perception of some customers was that they could use the product to read books, while others reported they could not read books with Kindle due to the bad screen quality, battery, or other reasons.

The star-ratings in fact reflect whether users' expectation is met or not. Inspired by this observation, here we use conjoint analysis to quantify the weight of different perceptions on reviewers' overall expectation for the product. We then categorize the affordances into the six categories of the Kano model based on the results of the conjoint analysis.

### 5.2 Conjoint analysis with the ordered logit model

We take each different review text as a conjoint-analysis survey response and the star rating, $R$, given by the reviewer as the reviewer's own choice. As star-rating is an ordinal discrete value, to estimate the weight of each perception mentioned in the review text to the star rating, we use ordered logit regression [36, 37]. The ordered logit model is based on the proportional odds assumption. It assumes that the coefficients that describe the relationship between the lowest value versus all higher values of the dependent variable are the same as those that describe the relationship between the next lowest value and all higher values. Conventionally, this assumption is tested by the significance of the parallel test (>0.05) [38].

---

[1] http://www.nltk.org/
[2] https://stanfordnlp.github.io/CoreNLP/
[3] https://spacy.io/
[4] https://wordnet.princeton.edu/
[5] https://github.com/cjhutto/vaderSentiment
[6] http://sentiwordnet.isti.cnr.it/
[7] https://www.god-helmet.com/wp/whissel-dictionary-of-affect/index.htm

The star-rating $R$ has five ordinal values: 1 star, 2 stars, 3 stars, 4 stars, and 5 stars. The logit model is therefore described by the following equations:

$$Pr(R = 5|X_i^{(1)}, X_i^{(2)})$$
$$= \frac{exp(\varepsilon_1 + \sum_i(\alpha_i X_i^{(1)} + \beta_i X_i^{(2)}))}{1 + exp(\varepsilon_1 + \sum_i(\alpha_i X_i^{(1)} + \beta_i X_i^{(2)}))}$$
$$Pr(R \geq 4|X_i^{(1)}, X_i^{(2)})$$
$$= \frac{exp(\varepsilon_2 + \sum_i(\alpha_i X_i^{(1)} + \beta_i X_i^{(2)}))}{1 + exp(\varepsilon_2 + \sum_i(\alpha_i X_i^{(1)} + \beta_i X_i^{(2)}))}$$
$$Pr(R \geq 3|X_i^{(1)}, X_i^{(2)}) \quad (1)$$
$$= \frac{exp(\varepsilon_3 + \sum_i(\alpha_i X_i^{(1)} + \beta_i X_i^{(2)}))}{1 + exp(\varepsilon_3 + \sum_i(\alpha_i X_i^{(1)} + \beta_i X_i^{(2)}))}$$
$$Pr(R \geq 2|X_i^{(1)}, X_i^{(2)})$$
$$= \frac{exp(\varepsilon_4 + \sum_i(\alpha_i X_i^{(1)} + \beta_i X_i^{(2)}))}{1 + exp(\varepsilon_4 + \sum_i(\alpha_i X_i^{(1)} + \beta_i X_i^{(2)}))}$$
$$Pr(R \geq 1|X_i^{(1)}, X_i^{(2)}) = 1$$

where $X_i^{(1)}$ and $X_i^{(2)}$ represent the opposite perceived quality that the reviews have on the $i$-th affordance $X_i$. Usually, $X_i^{(1)}$ denotes the *absence/non-existence* of the affordance, or relatively *low* affordance quality in human cognition, like "slow", "low", "traditional", etc., while $X_i^{(2)}$ denotes the *presence/existence* of the affordance, or relatively *high* affordance quality, like "fast", "high", "modern", etc. The value of $X_i^{(1)}$ and $X_i^{(2)}$ is binary: 0 or 1. $X_i^{(1)} = 1$ means that the reviewer perceived the quality of $X_i$ as relatively *low*, or $X_i$ is *absent*; $X_i^{(2)} = 1$ means that the reviewer perceived the quality of $X_i$ as relatively *high*, or $X_i$ is *existent*. Both $X_i^{(1)}$ and $X_i^{(2)} = 0$ means that the reviewer does not mention $X_i$, and he/she does not care about the quality of the affordance. $\alpha_i$ and $\beta_i$ denote the weights of the opposite perceived qualities of $X_i$ in the star rating. Their practical meaning can be explained by the following equation:

$$Ln\left(\frac{Pr_j}{1-Pr_j}\right) = \varepsilon_j + \sum_i(\alpha_i X_i^{(1)} + \beta X_i^{(2)}) \quad (2)$$

where $Pr_j = Pr(R \geq j|X_i^{(1)}, X_i^{(2)})$, and $j$ is the number of stars given by the reviewer. For example, when $X_i^{(1)}$ changes from 0 to 1, the odds of the reviewer giving more than j-star (i.e. higher star-rating) $\frac{Pr_j}{1-Pr_j}$ are multiplied by $exp(\alpha_i)$.

**5.3 Explaining the coefficients with the Kano model**

After $\alpha_i$ and $\beta_i$ are calculated, each pair of coefficients $\alpha_i$ and $\beta_i$ are plotted in the Cartesian coordinate system by two points: $A_i = (-1, \alpha_i)$ and $B_i = (1, \beta_i)$. As $X_i^{(1)} = 1$ mainly denotes the *absence* or the *low* quality of affordance, $\alpha_i < 0$ means that the *absence* (*low* quality) reduces the possibility of the reviewers giving a higher rating, whereas $\alpha_i > 0$ indicates that the *absence* (*low* quality) increases the possibility the reviewers giving a higher rating. The same holds for the coefficient $\beta_i$ and the *presence* (*high* quality) of the affordance $X_i^{(2)}$.

As illustrated in Figure 1, in the Kano model, the curves representing performance attribute and indifference attribute are relatively close to the origin (0, 0). The difference is that the performance attribute has a larger slope. The curves representing attractive attribute and must-be attribute are relatively far from the origin. The attractive attribute is situated above the horizontal axis and must-be attribute is situated below it. Based on this observation, we categorize the affordance $X_i$ in the Kano model based on the slope $K_i = \frac{\beta_i - \alpha_i}{2}$ and the intercept $M_i = \frac{\alpha_i + \beta_i}{2}$ of segment $A_i B_i$ (Figure 4) with the following rules (**Erreur ! Source du renvoi introuvable.**): if $K_i$ is negative, then the affordance $X_i$ is categorized as a reverse attribute. If $K_i$ is positive and is lower than the threshold $k$ ($k > 0$), if $-m \leq M_i \leq m$ ($m > 0$), then $X_i$ is categorized as an indifferent attribute. If $M_i > m$ or $M_i < -m$, $X_i$ is categorized as a questionable attribute. If $K_i$ is higher than the threshold $k$, $M_i > m$, $-m \leq M_i \leq m$ and $M_i < -m$ mean that $X_i$ is an attractive attribute, a performance attribute and a must-be attribute, respectively. The thresholds should be adequate to the objective of the task. For example, if the threshold $k$ is too high, then most affordances would be indifferent affordances. If the threshold $m$ is too low, then most affordances would be must-be or attractive affordances.

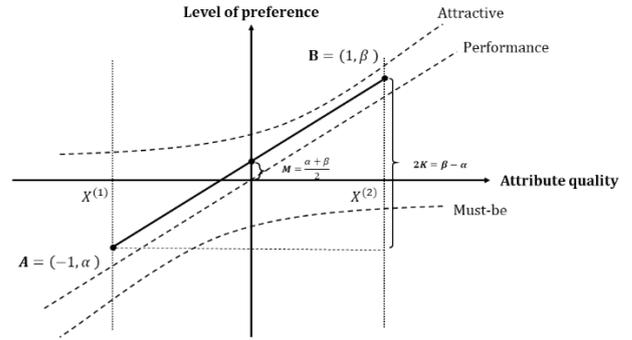

Figure 4 Parameters $K$ and $M$ illustrated on the Kano model

Table 1 Categorization rules according to the parameters $K$ and $M$ on the Kano model

| $K$ | $M$ | Categorization |
|---|---|---|
| $K < 0$ | | Reverse attribute |
| $0 < K < k$ | $M < -m$ or $M > m$ | Questionable attribute |
| | $-m < M < m$ | Indifferent attribute |
| $K > k$ | $M < -m$ | Must-be attribute |
| | $-m < M < m$ | Performance attribute |
| | $M > m$ | Attractive attribute |

The differences between our method of using the categories of the Kano model and the original Kano survey comes from the unstructured nature of online review data. First, the horizontal axis in the Kano model, originally represents the real quality (presence/absence) of the product attribute. In our utilization, we are analyzing users' perceived quality of the affordances. For example, it is known to all that an e-reader does provide readability. However, due to user incapability or user misuse, the perception of some reviewers is that they cannot read with it. Second, the responses in the Kano survey represent the absolute value of user preference level for the absence and presence of the attribute. However, in our study, the coefficients $\alpha_i$ and $\beta_i$ describe the odds of the reviewer giving a higher star-rating in cases where the reviewer mentions the *absence/presence* of the affordance ($X_i^{(1)} = 1$ or $X_i^{(2)} = 1$), compared with the case that the reviewer does not mention the *absence/presence* of the affordance ($X_i^{(1)} = 0$ or $X_i^{(2)} = 0$). Third, in a Kano survey, each participant is required to give his/her choices in two conditions, i.e. the *absence* of the attribute and the *presence* of the attribute, whereas in our study, as online review data is unstructured, reviewers do not have to mention every affordance of the product in their review text. Consequently, our method cannot be applied on individual reviewers. According to Arrow's impossibility theorem, it is not possible to convert ranked preferences of individuals into a consistent set of ranked preferences for the group. The categorization of affordance is therefore based on the *aggregated* expectation of the reviewer group. Finally, for "questionable attributes", they are ruled out of Kano surveys since people at the origin are considered as irrational. As in our case, numerous reviews are aggregated for a given affordance, this category may really appear as it only denotes that the opposite perceptions on the affordance divides people opinion.

### 5.4 Analyzing online reviews of different spans of time

By applying the proposed conjoint analysis method to the online reviews published in different time-spans, designers can observe the changes in the categorization of product affordances in the Kano model at different times.

## 6. Case study

Based on our discussion in Section 5, we demonstrate our proposed conjoint analysis method with the online review data of Kindle e-readers posted from 2013 to 2018. The online reviews are separated into two time-spans: from September 2013 to June 2015 and from July 2015 to April 2018. The objective is to observe the changes in user expectation in these two time-spans. In fact, two consecutively released versions of Kindle are involved in our case study: Kindle Paperwhite 2[1] (hereafter referred to as KP2) and Kindle Paperwhite 3[2] (hereafter referred to as KP3), which have similar market targets, as they were priced at the same level. KP2 was launched on September 2013 and was replaced by KP3 in July 2015 (Table 2). Therefore, the reviews of KP2 stands for the reviews posted in the first time-span, and the reviews of KP3 stands for the reviews posted in the second time-span. As is discussed in Section 5.4, online reviews can be collected from different products in the same product category.

Table 2 Product features of KP2 and KP3 and descriptive statistics of online review data

| Product name | On-shelf period | Price | Typical features | Number of reviews |
|---|---|---|---|---|
| Kindle Paperwhite 2 | Sep. 2013 - Jun. 2015 | Around $150 | Thickness:9.1mm; weight: 205g; screen: 212 ppi, 6 inches, 4 LEDs; battery: 8 weeks; storage: 4GBs | 45829 |
| Kindle Paperwhite 3 | Jul. 2015 – 2018 | Around $150 | Thickness:9.1mm; weight: 205g; screen: 300 ppi, 6 inches, 4 LEDs; battery: 6 weeks; storage: 4GBs | 56634 |

### 6.1 Data preparation

The data are prepared in the following steps. The statistics for each step are shown in Table 3. First, the credible reviews, which have at least one useful vote and are badged with verified purchase, are fed to our proposed rule-based natural language processing method to identify product affordances, usage contexts and the associated user perceptions. These elements are organized in the affordance description form (Section 4.1). Second, the authors carefully read the structured affordance descriptions that appear over a threshold (10 in our case study). The incorrect or unintelligible identification results are eliminated. Third, a frequently mentioned affordance is assumed to be more influential for the star rating. Therefore, the 50 most frequently occurring affordance descriptions for each product are chosen, which means that the proposed conjoint analysis is based on these 50 affordance descriptions. 30 of them have opposite perceptions and appear in both periods, which means we can observe dynamic changes in user expectation on these 30 affordances from 2013 to 2018.

Table 3 Descriptive statistics of the dataset.

| Steps | Statistics | KP2 | KP3 |
|---|---|---|---|
| Raw data | Nb. of reviews | 45829 | 56634 |
| Step 1 | Nb. of reviews selected | 8715 | 7922 |
| Step 1 | Nb. Of affordance descriptions extracted | 62681 | 60266 |
| Step 2 | Nb. of affordance descriptions extracted (appeared in more than 10 reviews) | 618 | 770 |
| Step 2 | Nb. of affordance descriptions extracted (after manual correction) | 565 | 680 |
| Step 3 | Nb. of affordance descriptions having opposite perceptions | 516 | 535 |
| Step 3 | Example of affordance descriptions having opposite perceptions (in the 50 most frequently appeared affordances) | read book, turn page, use kindle, buy kindle, use kindle, buy one, buy paperwhite, tell people, download book, buy this | read book, get one, use kindle, work kindle, make difference, find book, say that, try kindle, turn page |
| Step 3 | Nb. of affordance descriptions in common | 30 | |

### 6.2 Conjoint analysis results and representations on the Kano model

SPSS is used to calculate the coefficients $\alpha_i$ and $\beta_i$. In our case study, based on the conjoint analysis results, we choose the thresholds $k = 0.2$ and $m = 0.2$, which are adequate to observe the dynamic changes of the categorization of affordances over time. Table 4 illustrates the results of the conjoint analysis. 80% (96/120) of the coefficients are statistically significant. The signification in a parallel test for the KP2 and KP3 data is 0.054 and 0.105, respectively, which means the parallel assumption is validated (Section 5.2). Most of the opposite perceptions are non-existent and existent, only for connect WIFI-ability, and reviewers particularly perceive the speed of the connection, i.e. slow and fast.

Table 5 shows the categorization of affordances on the Kano model. Due to space limitations, the representation of the categorization results for ten affordances is illustrated in Figure 5.

As is mentioned, the scope of the concept of affordance is broad. What we are interested in is all the affordances that bring useful information for designers. For example, intuitively, the affordances buy *Kindle-ability* and *buy one-ability* do not involve physical interaction between the user and the product. Nevertheless, they describe a potential behavior that the customers can perform with the product. Meanwhile, one of the factors that is related to them is the price of the product. Only when the price is acceptable that the behavior "buy" can happen. That is why we keep this kind of affordance in our analysis.

For the affordance *travel a lot-ability*, we consider that it refers to the *travel-ability* with a kindle, as the underlying assumption of online review analysis is that the reviewers write their commentary related to the product.

For the affordance work kindle-ability, it refers to the ability for the Kindle to work correctly. Here, "kindle" should not an action receiver. Although it represents a mistake given by the open-sourced natural language processing package, we keep it because it is understandable for us.

From 2013 to 2015, ten affordances are categorized as must-be attributes, including as *work kindle-ability*, *turn page-*

---

[1] https://www.amazon.com/Amazon-Kindle-Paperwhite-eReader-Previous-Generation-6th/dp/B00AWH595M

[2] https://www.amazon.com/Amazon-Kindle-Paperwhite-6-Inch-4GB-eReader/dp/B00OQVZDJM

*ability*. Seven affordances are categorized as performance attributes, such as *read book-ability*, *change page-ability*. Three affordances are categorized as attractive attributes, such as *touch screen-ability*, *travel a lot-ability*. Eight affordances are categorized as indifferent attributes, such as *find book-ability*, *know word-ability*. *Return kindle-ability* is categorized as a reverse attribute and *try kindle-ability* is categorized as a questionable attribute. From 201 to 2018, fourteen affordances are categorized as must-be attributes, including *work kindle-ability*, *turn page-ability*. Four affordances are categorized as performance attributes, such as *read book-ability*, *take kindle-ability*. Seven affordances are categorized as indifferent attributes, such as *use kindle-ability*, *know word-ability*. Three affordances are categorized as reverse attributes, such as *upgrade kindle-ability*, *pay extra-ability*. Finally, *carry book-ability* is categorized as an attractive attribute, and *try kindle-ability* is always a questionable attribute.

Table 4 Estimated results of the parameters $\alpha$ and $\beta$[1]

| Affordance descriptions | Opposite perceptions ($p_1/p_2$) | KP2 (2013-2015) | | | | | | KP3 (2015-2018) | | | | | |
|---|---|---|---|---|---|---|---|---|---|---|---|---|---|
| | | α | Std. err | sig | β | Std. err | sig | α | Std. err | sig | β | Std. err | sig |
| read book | Non-existent/existent | -1.36 | 0.10 | ** | 1.02 | 0.05 | ** | -1.38 | 0.11 | ** | 0.99 | 0.05 | ** |
| get kindle | Non-existent/existent | -0.24 | 0.13 | ** | 0.00 | 0.07 | 0.33 | -0.19 | 0.12 | * | -0.11 | 0.06 | ** |
| use kindle | Non-existent/existent | -0.17 | 0.15 | * | 0.21 | 0.07 | ** | 0.01 | 0.13 | 0.30 | 0.12 | 0.06 | ** |
| work kindle | Non-existent/existent | -0.83 | 0.12 | ** | -0.11 | 0.08 | * | -0.85 | 0.13 | ** | -0.38 | 0.08 | ** |
| turn page | Non-existent/existent | -0.30 | 0.20 | ** | -0.19 | 0.08 | ** | -0.56 | 0.23 | * | -0.12 | 0.09 | ** |
| find book | Non-existent/existent | -0.18 | 0.16 | * | -0.19 | 0.09 | ** | -0.29 | 0.15 | ** | -0.17 | 0.08 | ** |
| know word | Non-existent/existent | 0.00 | 0.13 | 0.33 | 0.35 | 0.11 | * | -0.15 | 0.13 | * | 0.24 | 0.11 | * |
| try kindle | Non-existent/existent | -0.35 | 0.21 | ** | -0.21 | 0.09 | ** | -0.38 | 0.22 | ** | -0.29 | 0.09 | ** |
| buy kindle | Non-existent/existent | -0.91 | 0.22 | ** | 0.01 | 0.10 | 0.31 | -0.96 | 0.38 | ** | -0.08 | 0.17 | 0.22 |
| download book | Non-existent/existent | -0.78 | 0.25 | ** | 0.16 | 0.12 | * | -1.03 | 0.23 | ** | 0.17 | 0.11 | * |
| charge kindle | Non-existent/existent | -0.99 | 0.27 | ** | -0.24 | 0.12 | ** | -0.49 | 0.23 | ** | -0.30 | 0.12 | ** |
| upgrade kindle | Non-existent/existent | -0.88 | 0.20 | ** | -0.61 | 0.14 | ** | -0.62 | 0.22 | ** | -0.48 | 0.13 | ** |
| take kindle | Non-existent/existent | 0.12 | 0.43 | ** | 0.24 | 0.13 | * | -0.23 | 0.24 | * | 0.32 | 0.10 | ** |
| light screen | Non-existent/existent | 0.00 | 0.47 | 0.33 | 0.38 | 0.15 | ** | -0.80 | 0.57 | * | 0.36 | 0.16 | ** |
| read book at night | Non-existent/existent | -0.83 | 0.26 | * | 0.24 | 0.15 | * | -1.42 | 0.34 | 0.56 | -0.05 | 0.16 | 0.74 |
| buy one | Non-existent/existent | -0.55 | 0.30 | ** | 0.06 | 0.14 | 0.24 | -0.88 | 0.17 | ** | -0.03 | 0.08 | 0.25 |
| compare kindles | Non-existent/existent | -0.43 | 0.44 | * | 0.13 | 0.15 | * | -0.83 | 0.38 | 0.29 | -0.14 | 0.15 | * |
| change page | Non-existent/existent | 0.12 | 0.35 | 0.73 | 0.42 | 0.14 | ** | -0.30 | 0.30 | * | 0.12 | 0.13 | 0.28 |
| connect WIFI | Slow/fast | -0.65 | 0.34 | ** | -0.30 | 0.19 | * | -1.44 | 0.34 | ** | -0.29 | 0.18 | * |
| pay extra | Non-existent/existent | -0.26 | 0.34 | * | 0.15 | 0.17 | ** | -0.13 | 0.31 | * | -0.55 | 0.15 | ** |
| touch screen | Non-existent/existent | 0.19 | 0.35 | 0.20 | 0.69 | 0.15 | ** | -0.24 | 0.37 | 0.31 | -0.03 | 0.16 | * |
| add book | Non-existent/existent | -0.58 | 0.67 | 0.19 | 0.24 | 0.18 | * | -0.85 | 0.45 | * | 0.08 | 0.16 | 0.20 |
| travel a lot | Non-existent/existent | -0.08 | 0.51 | 0.29 | 0.79 | 0.19 | ** | -0.84 | 0.50 | ** | 1.10 | 0.20 | ** |
| own kindle | Non-existent/existent | -0.27 | 0.58 | ** | 0.08 | 0.20 | 0.71 | -0.20 | 0.37 | * | 0.17 | 0.18 | 0.05 |
| return kindle | Non-existent/Existent | -0.32 | 0.47 | * | -1.86 | 0.17 | ** | -0.03 | 0.33 | 0.31 | -1.55 | 0.12 | ** |
| leave charger | Non-existent/existent | -0.89 | 0.65 | * | -0.25 | 0.18 | * | -0.01 | 0.42 | 0.19 | -0.05 | 0.18 | 0.77 |
| carry book | Non-existent/existent | 0.73 | 1.08 | * | 1.56 | 0.25 | ** | 0.16 | 0.59 | ** | 0.29 | 0.19 | ** |
| adjust size | Non-existent/existent | -1.26 | 0.51 | ** | 0.92 | 0.21 | ** | -1.45 | 0.81 | ** | 0.99 | 0.19 | ** |
| replace kindle | Non-existent/existent | -0.36 | 0.57 | 0.18 | 0.18 | 0.18 | ** | -0.57 | 0.40 | * | -0.31 | 0.14 | ** |
| receive paperwhite | Non-existent/existent | -0.95 | 0.63 | ** | -0.17 | 0.21 | * | -0.67 | 0.48 | * | -0.17 | 0.18 | * |

Table 5 Categorization of affordance in the Kano model based on **K** and **M**[2]

| Affordance descriptions | Opposite perceptions ($p_1/p_2$) | KP2 | | | | | KP3 | | | | |
|---|---|---|---|---|---|---|---|---|---|---|---|
| | | α | β | K | M | Kano | α | β | K | M | Kano |
| read book | Non-existent/existent | -1.36 | 1.02 | 1.19 | -0.17 | P | -1.38 | 0.99 | 1.19 | -0.19 | P |
| get kindle | Non-existent/existent | -0.24 | 0.00 | 0.12 | -0.12 | I | -0.19 | -0.11 | 0.04 | -0.15 | I |
| use kindle | Non-existent/existent | -0.17 | 0.21 | 0.19 | 0.02 | I | 0.01 | 0.12 | 0.05 | 0.07 | I |
| work kindle | Non-existent/existent | -0.83 | -0.11 | 0.36 | -0.47 | M | -0.85 | -0.38 | 0.24 | -0.61 | M |
| turn page | Non-existent/existent | -0.30 | -0.10 | 0.10 | -0.20 | M | -0.56 | -0.12 | 0.22 | -0.34 | M |
| find book | Non-existent/existent | -0.18 | -0.19 | -0.01 | -0.19 | I | -0.45 | -0.02 | 0.22 | -0.24 | M |
| know word | Non-existent/existent | 0.00 | 0.35 | 0.17 | 0.18 | I | -0.15 | 0.24 | 0.20 | 0.04 | I |
| try kindle | Non-existent/existent | -0.35 | -0.21 | 0.07 | -0.28 | Q | -0.38 | -0.29 | 0.05 | -0.34 | Q |
| buy kindle | Non-existent/existent | -0.91 | 0.01 | 0.46 | -0.45 | M | -0.96 | -0.08 | 0.44 | -0.52 | M |
| download book | Non-existent/existent | -0.78 | 0.16 | 0.47 | -0.31 | M | -1.03 | 0.17 | 0.60 | -0.43 | M |
| charge kindle | Non-existent/existent | -0.99 | -0.24 | 0.38 | -0.61 | M | -0.25 | -0.04 | 0.11 | -0.15 | I |
| upgrade kindle | Non-existent/existent | -0.12 | 0.21 | 0.17 | 0.05 | I | -0.06 | -0.48 | -0.21 | -0.27 | R |
| take kindle | Non-existent/existent | 0.12 | 0.24 | 0.06 | 0.18 | I | -0.23 | 0.32 | 0.28 | 0.05 | P |
| light screen | Non-existent/existent | 0.00 | 0.38 | 0.19 | 0.19 | I | -0.80 | 0.36 | 0.58 | -0.22 | M |
| read book at night | Non-existent/existent | -0.83 | 0.24 | 0.54 | -0.30 | M | -1.42 | -0.05 | 0.68 | -0.74 | M |
| buy one | Non-existent/existent | -0.55 | 0.06 | 0.31 | -0.25 | M | -0.88 | -0.03 | 0.43 | -0.46 | M |
| compare kindles | Non-existent/existent | -0.43 | 0.13 | 0.28 | -0.15 | P | -0.83 | -0.14 | 0.35 | -0.48 | M |
| change page | Non-existent/existent | -0.12 | 0.42 | 0.27 | 0.15 | P | -0.30 | 0.12 | 0.21 | -0.09 | P |
| connect WIFI | Slow/fast | -0.65 | -0.30 | 0.18 | -0.47 | Q | -1.44 | -0.29 | 0.57 | -0.87 | M |
| pay extra | Non-existent/existent | -0.26 | 0.15 | 0.21 | -0.06 | P | -0.13 | -0.55 | -0.21 | -0.34 | R |
| touch screen | Non-existent/existent | 0.19 | 0.69 | 0.25 | 0.44 | A | -0.24 | -0.03 | 0.11 | -0.14 | I |
| add book | Non-existent/existent | -0.58 | 0.24 | 0.41 | -0.17 | P | -0.85 | 0.08 | 0.47 | -0.38 | M |
| travel lot | Non-existent/existent | -0.08 | 0.79 | 0.43 | 0.36 | A | -0.84 | 1.10 | 0.97 | 0.13 | P |
| own kindle | Non-existent/existent | -0.27 | 0.08 | 0.17 | -0.10 | I | -0.20 | 0.17 | 0.19 | -0.02 | I |
| return kindle | Non-existent/Existent | -0.32 | -1.86 | -0.77 | -1.09 | R | -0.03 | -1.55 | -0.76 | -0.79 | R |
| leave charger | Non-existent/existent | -0.89 | -0.25 | 0.32 | -0.57 | M | -0.05 | -0.01 | 0.02 | -0.03 | I |
| carry book | Non-existent/existent | 0.73 | 1.56 | 0.42 | 1.15 | A | 0.16 | 0.29 | 0.07 | 0.23 | A |
| adjust size | Non-existent/existent | -1.26 | 0.92 | 1.09 | -0.17 | P | -1.45 | 0.99 | 1.22 | -0.23 | M |
| replace kindle | Non-existent/existent | -0.36 | 0.18 | 0.27 | -0.09 | P | -0.57 | -0.13 | 0.22 | -0.35 | M |
| receive paperwhite | Non-existent/existent | -0.95 | -0.17 | 0.39 | -0.56 | M | -0.67 | -0.17 | 0.25 | -0.42 | M |

---

[1] For KP2, $R^2=0.0908$, sig = 0.054, and for KP3, $R^2=0.1069$, sig=0.105. Significance level: **, * are statistically significant at the 0.01 and 0.05 level, respectively.

[2] "P" means "performance attribute", "I" means "indifferent attribute", "M" means "must-be attribute", "A" means "attractive attribute", "R" means "reverse attribute", "Q" means "questionable attribute"

## 6.3 Analysis of the results and product improvement strategies

Kano [27] observed that in the Kano model, product attributes should appear as attractive and evolve towards must-be. Globally, 25 out of 30 affordances follow this observation.

For the affordances that do not change their categorization in our analysis results, *read book-ability* and *change page-ability* have always been performance attributes from 2013 to now. It is obvious that an e-reader with good readability constantly provides high-level customer loyalty (Section 2.5). Note, however, that unlike *read book-ability*, the presence of *read book at night-ability* does not produce much satisfaction, which suggests that improving *read book-ability* in other usage contexts may have a more positive influence on user satisfaction.

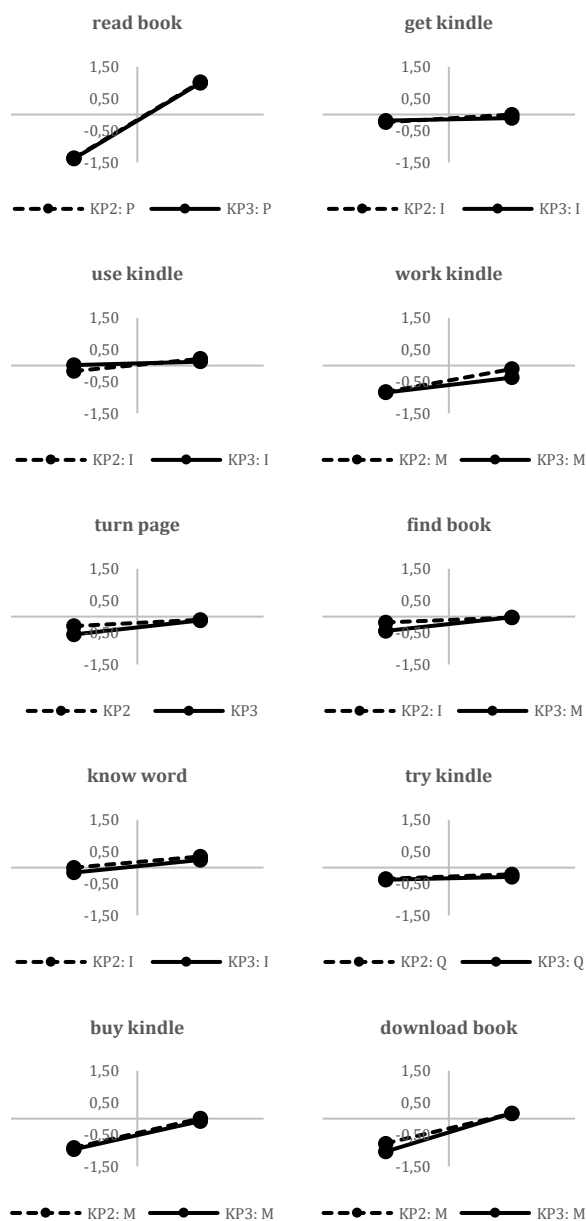

Figure 5 Representation of KP2 and KP3 product affordances on the Kano model

*Get kindle-ability*, *use kindle-ability*, *own kindle-ability* are constantly categorized as indifferent affordances. We assume the reason is that these affordances are too general in meaning. User expectation on these affordances is randomly distributed. For example, for *use kindle-ability*, people may use the Kindle to read, or to do other things. Literally, *get kindle-ability* refers to the act of buying or receiving. However, when we read the online reviews containing "get kindle", we find that the complements in the sentence may change completely its meaning. For example, "get the kindle upgraded" or "get it repaired". Due to the multiplicity of the meaning of the word "get", it is reasonable that its categorization is "indifferent affordance". *Know word-ability* is categorized as an indifferent affordance, which means customers pay less attention to it. Therefore, the implementation of a dictionary in the operating system is not essential.

*Work kindle-ability*, *turn page-ability*, *buy kindle-ability*, *download book-ability*, *read book at night-ability*, *buy one-ability*, *connect WIFI-ability*, and *receive paperwhite-ability* are constantly categorized as must-be affordances for both products. *Buy kindle-ability* and *buy one-ability* are synonymous affordances, so it is reasonable for them to be categorized in the same group.

Only *carry book-ability* remains an attractive affordance. However, it is much less "attractive" from 2015 to 2018. *Try kindle*-ability is always a questionable attribute. This means that customers get unsatisfied whether they try kindle or not before purchase. We find that in the online reviews, when reviewers talk about try kindle, they either express their regret for not having tried the e-reader at the store or tend to criticize the difference between the e-reader they had tried in the store and the e-reader they had received.

For the affordances that change categories, unsurprisingly, *travel lot-ability* changed from an attractive attribute to a performance attribute. *Compare kindles-ability*, *add book-ability*, *adjust size-ability*, and *replace kindle-ability* changed from performance attributes to must-be attributes. *Find book-ability* and *light screen-ability* turned from indifferent attributes to must-be attributes. *Take kindle-ability* changed from indifferent attribute to performance attribute. These trends support the study of Kano [27].

Interestingly, we found that *upgrade kindle-ability* was an indifferent attribute that is fast becoming a reverse attribute. In fact, according to Amazon's marketing strategy, each version of the Kindle e-reader is sold in two different configurations: one with advertisements and one without advertisements. The cheaper one constantly shows advertisements on the e-reader home screen. From the year 2014, customers have the option to upgrade kindle by paying an extra 20 dollars to stop getting advertisements. From 2013 to 2015, this was an attractive option, which means that customers are satisfied if they can upgrade the kindle. However, since 2015, customers are voicing dissatisfaction even if they can remove the advertising. We read the reviewers concerning this affordance, and we found that today's customers are tired of this marketing strategy. They reported that the upgrade option is just a trick to make them pay more money. This observation is supported by its synonymous affordance *pay extra-ability*, which shifts from a performance attribute to a reverse attribute.

Meanwhile, we observe that *charge-kindle ability* tends to become an indifferent affordance as the parameter $M$ gets higher. Our assumption is that compared with today's other electronic products, like smartphones, e-readers have a much larger battery capacity for ordinary use (i.e. about one month). However, it is also getting easier to find Kindle Paperwhite-compatible battery chargers as the micro-USB connector is becoming increasingly common on electronic products. This assumption is supported by its synonymous affordance *leave charger-ability*, which is also changing from a must-be attribute to an indifferent attribute. This means that from 2013 to 2015, if users cannot/do not leave the charger at home or at other places that they used to go to, then they are unsatisfied. However, from 2015 to 2018, charger availability is less of an issue for users.

The move from KP2 to KP3 marked an increase in screen resolution and a decrease in battery capacity (Table 2). As *read book-ability* remains an important performance attribute while *charge kindle-ability* is becoming less of a must-be attribute, these upgrades respond to the dynamic changes in user expectation found in our analysis. Our study suggests that for next-generation e-readers, designers should pay less attention to battery and storage capacity, and more attention to their market strategy. Selling the *with advertisements-version* is a questionable strategy. Also, *read book-ability* in general is a performance attribute, while *read book at night-ability* is a must-be attribute, which suggests that improving reading experience in other usage contexts—such as reading *in the sun*, *on plane*, *on the beach,* for example—may help improve user satisfaction.

## 7. Discussion and Conclusions

Online reviews have been studied by many researchers in product design due to their rich content and high reliability. Our analysis is focused on how to follow the changes in user expectation. To accomplish this, we combine affordance-based design, natural language processing, conjoint analysis and Kano modeling in a unique manner. Our work has useful implications for data-driven design.

### 7.1 Theoretical implications

First, today's big trend is big data analytics [39]. However, compared with traditional data, if nothing new can be discovered from big data, then why bother with big data analytics? Through our study, we find that the value of online review data added to product design depends on its unprecedented characteristic, which is critical to creating actionable new insights for decision-making. That is where data analytics should begin.

Second, we have to use the right domain theory to change the unstructured text data to structured data before further analysis. Research to date has been dominated by feature-based opinion mining methods, which involve extracting product features, extracting perceptual words, and determining sentiment orientation. However, the concept of product feature lacks a theoretical basis in design engineering. As previous research found, it cannot cover all the significant issues addressed in customer reviews for product design. Users are not focused just on product features, but also on the usage and usage contexts of the product, which correspond to the concept of affordance in affordance-based design. That is why we introduce the concept of affordance to structure the text data.

Third, Qi, Zhang, Jeon and Zhou [15] insisted that the classical design models should be reformed in the context of online review data. Our research supports Qi et al.'s opinion. The Kano model, for example, has been widely used in product development for many years. Kano model analysis has always been based on physical prototypes and focus groups. In our study, we reform the model in the context of online review data. Another example is the affordance description form, which we have revisited by adding usage context and perceived quality in order to match reviewers' linguistic patterns.

### 7.2 Practical implications

Online reviews provide large amounts of data to for mining mine user requirements and preferences. Our research provides an approach to process data structuring and data analytics. In particular, a conjoint analysis method is proposed to quantitatively categorize the automatically-structured affordances into the Kano model. We demonstrated with a case study that using our proposed method, designers are able to find unexpected changes in user expectation for product affordances. It is thus practical to evaluate the improvement strategies in previous generations of product, and to propose new strategies for designing the next generation of the product. Our approach can be easily and usefully extended in various industries for different kinds of popular products, from mobile phones and wearable devices to electrical household appliances.

### 7.3 Limitations

First, our research provides a pioneering study on how to extract and structure affordance from online reviews in a highly automated manner. However, the performance of the proposed rule-based natural language processing method still requires human effort to eliminate the errors in identification results. The performance can be improved by adding more linguistic rules to the method and employing more accurate natural language processing algorithms.

Second, our analysis of the dynamic changes in user expectation in Section 6.3 is mainly based on our revisions of the online reviews concerning the related affordances. Future research is needed on how to combine the present analysis of anonymous online review data and nominative data provided by conventional user requirements identification methods like surveys, questionnaires and focus groups, in order to better understand the dynamic changes in user expectation and validate the product improvement strategies.

Third, our research reveals the possibility of monitoring the changes in user expectations in short periods, like 6 months. However, a further experiment needs to be done to validate this assumption.